\documentclass[11pt]{article}
\usepackage{acl}
\usepackage[ruled,vlined]{algorithm2e}
\usepackage{amsfonts}
\usepackage{amsmath}
\usepackage{amssymb}
\usepackage{amsthm}
\usepackage{bm}
\usepackage{booktabs}
\usepackage[T1]{fontenc}
\usepackage{graphicx}
\usepackage{inconsolata}
\usepackage[utf8]{inputenc}
\usepackage{latexsym}
\usepackage{makecell}
\usepackage{microtype}
\usepackage{multirow}
\usepackage{subcaption}
\usepackage[most]{tcolorbox}
\tcbset{
  promptbox/.style={
    enhanced,
    breakable,
    colback=white,
    colframe=black!40,
    boxrule=0.5pt,
    arc=2pt,
    left=6pt,right=6pt,top=6pt,bottom=6pt,
    fonttitle=\bfseries,
    title={#1},
  }
}
\usepackage{times}

\newcommand{\takehiro}[1]{\textcolor{black}{#1}}

\title{From Heard to Lived Opinions:\\ Simulating Opinion Dynamics with Grounded LLM Agents\\ in Economic Environments}

\author{
 \textbf{Ryuji Hashimoto\textsuperscript{1,2}},
 \textbf{Masahiro Kaneko\textsuperscript{1,3}},
 \textbf{Ryosuke Takata\textsuperscript{1,2}},
 \textbf{Takehiro Takayanagi\textsuperscript{1,2}},
 \textbf{Kiyoshi Izumi\textsuperscript{1,2}},
 \\
\textsuperscript{1}Simulacra Inc.,
\textsuperscript{2}The University of Tokyo,
\textsuperscript{3}MBZUAI
\\
\small{
    \textbf{Correspondence:} \href{mailto:email@domain}{r\_hashimoto@simulacra.co.jp}
}
} 

\begin{document}
\maketitle
\begin{abstract}
\takehiro{
Opinion dynamics (OD) studies how individual opinions evolve and generate collective patterns such as consensus and polarization. While recent work explores OD using populations of LLM-based agents focusing on opinion exchange, it typically does not incorporate individuals’ lived experiences, such as economic outcomes of past decisions, which play a critical role in shaping opinions. We propose a novel OD simulation framework that grounds LLM-based agents in an economic environment, allowing them to act and receive environmental feedback. Our simulations exhibit coherent OD at both individual and population levels: individual opinions follow structured trajectories shaped by economic experiences, with adverse conditions inducing opinion rigidity, while at the population level, collective opinions co-move with economic conditions, with inequality amplifying polarization and price instability driving larger distributional shifts. These results highlight the importance of grounding LLM-based agents in environments to capture collective OD.
}
\end{abstract}

\section{Introduction}
Research in opinion dynamics (OD) examines how individual opinions form and update through interactions, giving rise to collective patterns~\citep{degroot_model,hegselmann_krause_model}. Collective opinion distributions are not mere aggregations of individual opinions; rather, they emerge through nonlinear interactions among individuals~\citep{od_review}. Because such dynamics influence political decision-making and the stability of economic and social institutions, understanding opinion formation and distribution is fundamental to social analysis~\citep{od_importance}.

Recently, large language models (LLMs) have been applied to OD simulations~\citep{llm_fallacy,bias_llm_opinions1,bias_llm_opinions2} due to their ability to update opinions through language–level semantic understanding, to represent multifaceted opinions without predefining topics, and to naturally capture human-like values. Prior studies demonstrate that LLM-based OD are shaped by confirmation bias and framing effects~\citep{bias_llm_opinions1}, as well as by the content and style of dialogue~\citep{llm_fallacy}.


\takehiro{However, most existing OD studies focus primarily on opinion exchange, with opinion change driven largely by linguistic interaction. In contrast, real-world opinion formation is also shaped by lived experiences, such as economic actions and their consequences, which are largely absent from language-only simulations.}
For example, exposure to economic threat can induce rigidity in attitudes~\citep{threat_rigidity}. Likewise, economic inequality has been shown to foster polarization~\citep{us_gini_pol}. Prior work lacks an explicit environment in which agents’ actions generate economic experiences and feedback; as a result, phenomena such as threat-induced rigidity or inequality-driven polarization cannot be modeled as endogenous outcomes of interaction dynamics.

\takehiro{To fill this gap, we propose a novel OD simulation framework in which LLM-based agents are grounded within an economic environment and update their opinions through feedback arising from their own actions and experiences.} Specifically, LLM-based agents are modeled as households that make economic decisions while also exchanging opinions about the society. These decisions endogenously affect the environment, leading to changes in prices, wages, and asset positions, which constitute agents’ lived experiences. The accumulated experiences, in turn, shape subsequent decision-making and opinion formation.

\takehiro{To demonstrate that grounding LLM-based agents in an economic environment enables the observation of OD driven by actions and environmental feedback, we examine whether this approach produces coherent OD at both the individual and population levels. To this end, we investigate a sequence of research questions.}
\paragraph{\textbf{RQ1}: \takehiro{Do LLM-based agents grounded in an economic environment exhibit environment-consistent actions and opinions?}}
\takehiro{We first assess the basic validity of the framework at the individual level by examining whether agents’ actions are appropriate to their environment and whether differences in experience lead to plausible qualitative differences in expressed opinions.}
\paragraph{\textbf{RQ2}: How do histories of action and feedback shape individual OD?}
\takehiro{Then, we examine how each agent's history of actions and environmental feedback influence the way their opinions evolve over time. This allows us to characterize individual OD as experience-dependent processes.}
\paragraph{\takehiro{\textbf{RQ3}: How do individual-level economic interactions scale up to population-level OD?}}
\takehiro{Finally, we analyze how population-level opinion distributions co-evolve with macroeconomic dynamics, focusing on the relationship between economic inequality and opinion polarization, as well as between macroeconomic instability and temporal instability in opinion distributions.}
Our results show that introducing an economic environment in LLM-based OD simulations makes OD grounded, endogenous, and experience-driven. At the individual level, agents exhibit economically plausible behavior and experience-dependent opinion trajectories, including increased rigidity under macroeconomic instability. At the population level, opinion distributions co-evolve with economic conditions, with greater inequality associated with polarization and higher price volatility associated with greater opinion instability.


\section{\takehiro{LLM-based OD Simulation with Economic Environments}}

\begin{figure}[tbp]
  \centering
  \includegraphics[width=\linewidth]{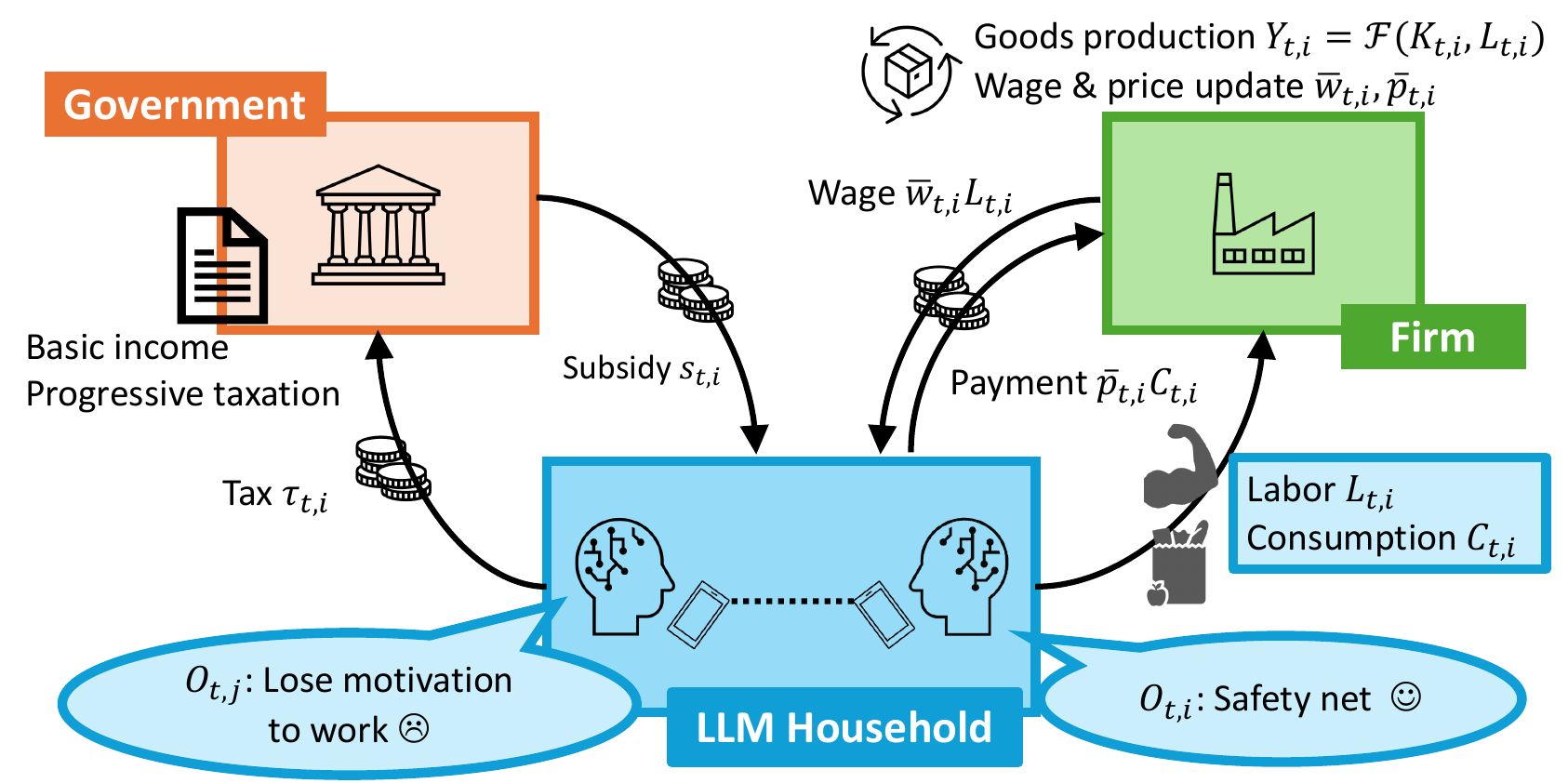}
  \caption{Structure of the LLM-based OD simulations. At each time step $t\in\{1,...,T\}$, LLM-based household $i\in\{1,...,n\}$ sequentially makes economic decisions regarding labor supply $L_{t,i}$ and consumption $C_{t,i}$. A rule-based firm produces goods $Y_{t,i}$ using household's labor and sets wage $\bar{w}_{t,i}$ and price $\bar{p}_{t,i}$ according to demand-supply imbalance. A government levies taxes $\tau_{t,i}$ and provides subsidies $s_{t,i}$ to households. In parallel, LLM-based agents exchange their opinions $O_{t,i}$ about the societal and economic environment.}
  \label{Fig:simulation_structure}
\end{figure}

Assume $n\in\mathbb{N}$ LLM-based agents operate in a single macroeconomic environment setting, with each simulation consisting of $T\in\mathbb{N}$ time steps. As illustrated in Figure~\ref{Fig:simulation_structure}
, the LLM-based agents are modeled as households, while the firm and the government are implemented as rule-based agents. At each time step $t\in\{1,\ldots,T\}$, household agents are iterated over sequentially. After each household $i\in\{1,\ldots,n\}$ makes its labor, consumption, and opinion decisions, the macroeconomic environment---comprising the firm and the government---transitions according to rule-based update rules, and the updated state is observed by the subsequent household agent. In the following, we detail the sequence of operations at time step $t$ during the turn of household $i$, describing in order the actions of the household, firm, and government.

\paragraph{\textbf{LLM Household}} During the decision turn of household $i$ at time step $t$, the LLM outputs categorical actions $a^L_{t,i}$ and $a^C_{t,i}$, where $a^L_{t,i}, a^C_{t,i} \in \{\text{low}, \text{medium}, \text{high}\}$, together with a natural-language opinion $O_{t,i}$.
The categorical actions are subsequently mapped to numerical values via a post-processing step. For labor and consumption, we define continuous intervals $\mathcal{I}^L_{\cdot}$ and $\mathcal{I}^C_{\cdot}$, respectively. The realized labor $L_{t,i}$ and consumption $C_{t,i}$ are obtained by uniform sampling from the intervals associated with the selected categories:
{\normalsize\begin{align}
L_{t,i} \sim \mathcal{U}(\mathcal{I}^L_{a^L_{t,i}}), \qquad C_{t,i} \sim \mathcal{U}(\mathcal{I}^C_{a^C_{t,i}})
\end{align}}
The interval boundaries are predefined and remain constant throughout the simulation. After completing its decision-making at stage $(t, i)$, the household’s cash holding $M_{t,i}\in\mathbb{R}$ is updated according to the labor supply and consumption choices as: 
{\normalsize\begin{align}
M_{t,i}\leftarrow M_{t-1,i}-\bar{p}_{t,i}C_{t,i}+\bar{w}_{t,i}L_{t,i}\label{Eq:household_income}
\end{align}}
where $\bar{p}_{t,i}$ and $\bar{w}_{t,i}$ denotes the price per unit of goods and wage per unit of labor.

The LLM’s outputs are conditioned on a structured prompt. The prompt consists of: (i) a role instruction specifying the household’s decision task; (ii) a persona description for role-play; (iii) public information such as historical wages, prices, taxes and subsidies; (iv) observed opinions of other households; (v) the household’s own financial history; and (vi) the household’s own opinion expressed in the previous time step. The household’s financial history is summarized in terms of its relative financial status within the population. Using the mean $\mu_t$ and standard deviation $\sigma_t$ of all households at time $t$, the relative financial position of household $i$ is measured by its standardized deviation from the mean, $z_{t,i} = (M_{t,i} - \mu_t) / \sigma_t$. The household’s financial status $f_{t,i}$ takes one of five categories: very high, above-average, around-average, below-average or very low, determined according to predefined thresholds on $z_{t,i}$.

\paragraph{\textbf{Firm}} Following the household decision-making described above, the macroeconomic environment is updated through the actions of a rule-based firm. The firm updates production, prices, and wages based on the current intermediate state of the economy. The firm produces a homogeneous consumption good using agent $i$'s labor at time $t$ and capital. Production follows a Cobb–Douglas technology function~\citep{cobb_douglas}. Let $K_{t,i}$ denote the firm’s effective capital stock at the beginning of stage $(t,i)$. Production of goods $Y_{t,i}=\mathcal{F}(K_{t,i},L_{t,i})$ is given by
{\normalsize\begin{align}
\mathcal{F}(K_{t,i},L_{t,i}) = A (\min(K_{\text{max}},K_{t,i}))^\alpha  (L_{t,i})^{(1 - \alpha)}
\end{align}}
where $A > 0$ is the total factor productivity, $\alpha \in (0,1)$ is the capital share, and $K_{\text{max}}$ is the maximum amount of capital used for one-time production. The produced output is consumed by the household and added to the firm’s inventory:
{\normalsize\begin{align}
K_{t,i+1}\leftarrow (1-d)K_{t,i}+Y_{t,i}-C_{t,i}\label{Eq:update_inventory}
\end{align}}
, which depreciate between stages with rate $d\in(0,1)$. After observing household $i$'s demand $C_{t,i}$ and supply $Y_{t,i}$ at the current stage, the firm updates the price of unit of the good $\bar{p}_{t,i}$. The price is adjusted according to the imbalance between demand $C_{t,i}$ and supply $Y_{t,i}$:
{\normalsize\begin{align}
\bar{p}_{t,i+1}\leftarrow \bar{p}_{t,i} \left(1+\eta_p \frac{C_{t,i}-Y_{t,i}}{Y_{t,i}}\right)\label{Eq:update_price}
\end{align}}
where $\eta_p$ is a price elasticity.

After providing wage to the household $\bar{w}_{t,i}L_{t,i}$, the firm updates the wage per unit of labor $\bar{w}_{t,i}$ in response to productivity conditions implied by the current production state. The wage is adjusted toward the marginal product of labor $\text{MPL}_{t,i}=\partial Y_{t,i}/\partial L_{t,i}$ using a wage elasticity $\eta_w > 0$:
{\normalsize\begin{align}
\bar{w}_{t,i+1}&\leftarrow\bar{w}_{t,i}+\eta_w(\text{MPL}_{t,i}-\bar{w}_{t,i})\label{Eq:update_wage}\\
\text{MPL}_{t,i}&=(1-\alpha)A(\min(K_{\text{max}},K_{t,i}))^\alpha L_{t,i}^{-\alpha}
\end{align}}

\paragraph{\textbf{Government}} Following the firm update, a rule-based government implements a simplified basic income and progressive taxation scheme. The government publicly announces the current policy to households at each stage $(t, i)$. Each household receives a fixed amount of basic income as subsidies: $\forall (t,i)~~ s_{t,i}=b$. The basic income payment is unconditional and does not depend on the household’s current labor supply or consumption choices. Also, the government levies a tax that depends on the household’s relative financial status. $\bar{\tau} \in [0,1)$ denote the base tax rate, which serves as the reference rate for taxation. For household $i$ at stage $(t, i)$, the effective tax $\tau_{t,i}$ is determined as
{\normalsize\begin{align}
\tau_{t,i}=\begin{cases}
0 \quad\quad~~~ \text{if}~~ f_{t,i}=\text{very low}\\
2\bar{\tau}M_{t,i} ~~ \text{if}~~ f_{t,i}=\text{very high}\\
\bar{\tau}M_{t,i}~~~~\text{otherwise}
\end{cases}
\end{align}}
As a result, after stage $(t,i)$, the household's income is updated as $M_{t,i}\leftarrow M_{t,i}+s_{t,i}-\tau_{t,i}$.

\section{Experimental Settings}
\subsection{Simulation Configuration} 
We run 30 trials of simulations with $T=150$ and $n=20$. \takehiro{LLM-based agents are implemented using Llama 3.1 8B~\cite{llama3} with a sampling temperature of 0.7.}
For each agent, a persona prompt is constructed using demographic attributes drawn from a publicly available persona dataset~\citep{nvidia_personas_usa} and personality traits based on the Big Five personality model~\citep{big_five}. All other hyperparameters, as well as detailed simulation procedure and computing infrastructure are provided in Appendices~\ref{Sec:hyperparameter} and \ref{Sec:computing_infrastructure}.

\subsection{Data Collection and Processing} We remove the first 10 time steps ($t\leq10$) of each simulation run to eliminate transient effects arising from the initial conditions. At the end of each stage $(t,i)$, we record all variables including agent's actions $L_{t,i},C_{t,i}$, opinion $O_{t,i}$, and environmental contexts $\bar{w}_{t,i},\bar{p}_{t,i},f_{t,i}$. In addition, we compute and record the time-$t$ averages of wages and prices, $\bar{w}_t$ and $\bar{p}_t$. The textual opinion $O_{t,i}$ is further transformed into a numerical embedding $\bm{o}_{t,i}\in\mathbb{R}^{384}$ using pretrained sentence transformer~\citep{minilm}, and a scalar sentiment score $\psi_{t,i}\in[-1,1]$ using pretrained sentiment analysis model~\citep{finbert}. The sentiment score provides a normalized measure of the agent’s expressed attitude, where $-1$ and $+1$ correspond to maximally negative and positive sentiment, respectively. 

\paragraph{\textbf{Individual States}} Agents in our simulation exhibit heterogeneous and continuously evolving economic conditions and generate free-form textual opinions. Directly analyzing such high-dimensional and unstructured information makes it difficult to identify common patterns in opinion formation. To address this challenge, we introduce a compact representation of an agent’s situation at each time step, termed the {\em internal state}. To characterize agents’ internal states in a structured manner, we discretize both expressed sentiment and financial status into coarse-grained categories and record their joint configuration at each stage $(t,i)$. First, the continuous sentiment score $\psi_{t,i}$ is mapped into three sentiment classes using fixed threshold $\theta^\psi$:
{\normalsize\begin{align}
\tilde{\psi}_{t,i}=\begin{cases}
\text{NEG},\quad \psi_{t,i}<-\theta^{\psi}\\
\text{NEU},\quad -\theta^{\psi}\leq\psi_{t,i}\leq\theta^{\psi}\\
\text{POS},\quad \theta^{\psi}<\psi_{t,i}
\end{cases}
\end{align}}
where $\text{NEG}$, $\text{NEU}$, and $\text{POS}$ mean negative, neutral, and positive, respectively. Using this representation, we can analyze the sentiment change rates, defined as the fraction of agents whose sentiment class changes between consecutive time steps:
{\normalsize\begin{align}
\delta_t&=\frac{1}{n}\sum_i \textbf{1}(\tilde{\psi}_{t,i}\neq\tilde{\psi}_{t-1,i}),\\
\delta_t^{'}&=\frac{1}{n}\sum_i\textbf{1}(\tilde{\psi}_{t,i}\neq\tilde{\psi}_{t-1,i}\land\tilde{\psi}_{t,i}\neq\text{NEU})
\end{align}}
Next, the agent's financial status $f_{t,i}$, originally drawn from five qualitative categories, is coarsened into three aggregate levels:
{\normalsize\begin{align}
\tilde{f}_{t,i}=\begin{cases}
\text{low},\quad~~~~ f_{t,i}\in\{\text{very low},\text{below-average}\}\\
\text{average},~ f_{t,i}\in\{\text{around-average}\}\\
\text{high},\quad~~~ f_{t,i}\in\{\text{very high},\text{above-average}\}
\end{cases}
\end{align}\label{Eq:tilde_f}}
We then define the state of agent $i$ at time $t$ as the Cartesian product of sentiment and economic categories:
{\normalsize\begin{align}
h_{t,i}= (\tilde{\psi}_{t,i}, \tilde{f}_{t,i}) \in\mathcal{H}
\end{align}}
where $\mathcal{H}$ denotes a finite internal state space:
{\normalsize\begin{align}
\mathcal{H}= \{\text{NEG}, \text{NEU}, \text{POS}\}\times\{\text{low}, \text{average}, \text{high}\}
\end{align}}

Furthermore, we record the temporal transitions of internal states for each individual, $h_{t-1,i}\rightarrow h_{t,i}$, throughout the simulation. As all internal state transitions are represented by ordered pairs $(h,h')\in\mathcal{H}\times\mathcal{H}$, the total number of distinct transition types is $9\times9=81$. For each individual $i$, we define the number of occurrences of a given transition $(h,h')$ over the entire simulation as
{\normalsize\begin{align}
N_i(h,h')=\sum_t\textbf{1}(h_{t-1,i}=h,h_{t,i}=h')
\end{align}}
where $\textbf{1}(\cdot)$ denotes the indicator function. By arranging the counts of all 81 transition types in a fixed order, we obtain a transition feature vector $\bm{n}_i\in\mathbb{N}^{81}$ for individual $i$,
{\normalsize\begin{align}
\bm{n}_i={}^\top\!\begin{pmatrix}N_i(h_1,h_1) & \ldots& N_i(h_9,h_9)\end{pmatrix}
\end{align}}
where $\{h_1,...,h_9\}$ denotes an arbitrary but fixed ordering of the elements of $\mathcal{H}$. 

By considering a finite internal state space, we can 1)abstract away individual-level variability, making it easier to identify representative behavioral patterns, and 2)gain structured link between agents’ economic experiences and their expressed opinions. This representation enables us to explicitly trace how individual economic experiences shape opinion and, in turn, drive the dynamics of population-level opinion distributions, which are examined in RQ1–RQ3. For example, In RQ2, we apply KMeans clustering to the collection of state-transition count vectors $\{\bm{n}_i\}$ across all agents and simulation runs, grouping individuals according to the similarity of their OD transition patterns.

\paragraph{\textbf{Population-level descriptors}} To assess internal states and trajectories as collective phenomena, we next introduce population-level descriptors that summarize the collective state of the system at each time step $t$. First, to quantify the degree of economic inequality in the population, we define Gini coefficient at time $t$ as:
{\normalsize\begin{align}
G_t=\frac{1}{2n^2\bar{M}_t}\sum_{i=1}^{n}\sum_{j=1}^{n}\left| M_{t,i} - M_{t,j} \right|
\end{align}}
where $\bar{M}_t$ denotes the average income. 

To characterize population-level OD, we consider two-types of opinion distributions---sentiment scores $\Psi_t=\{\psi_{t,i}\}_{i=1}^n$ and opinion embeddings $\mathcal{O}_t=\{\bm{o}_{t,i}\}_{i=1}^n$. To complement the standard deviation of sentiment as an indicator of opinion dispersion, we define a polarization index $P_t^\Delta$, which is defined based on the difference between the mean sentiment of positive and negative group, while explicitly incorporating group-size balance:
{\normalsize\begin{align}
P_t^\Delta(\theta^\psi)&=4|\mu_t^{\text{POS}}-\mu_t^{\text{NEG}}|r_t^{\text{POS}}r_t^{\text{NEG}}\\
\mu_t^{\tilde{\psi}}&= \frac{1}{{n_t^{\tilde{\psi}}}}\sum_i\textbf{1}(\tilde{\psi}_{t,i}=\tilde{\psi})\psi_{t,i}\\
r_t^{\tilde{\psi}}&= \frac{n_t^{\tilde{\psi}}}{n},~~n_t^{\tilde{\psi}}=\sum_i\textbf{1}(\tilde{\psi}_{t,i}=\tilde{\psi})
\end{align}}

\section{Results and Discussions}


\begin{figure}[tbp]
  \centering
  \includegraphics[width=\linewidth]{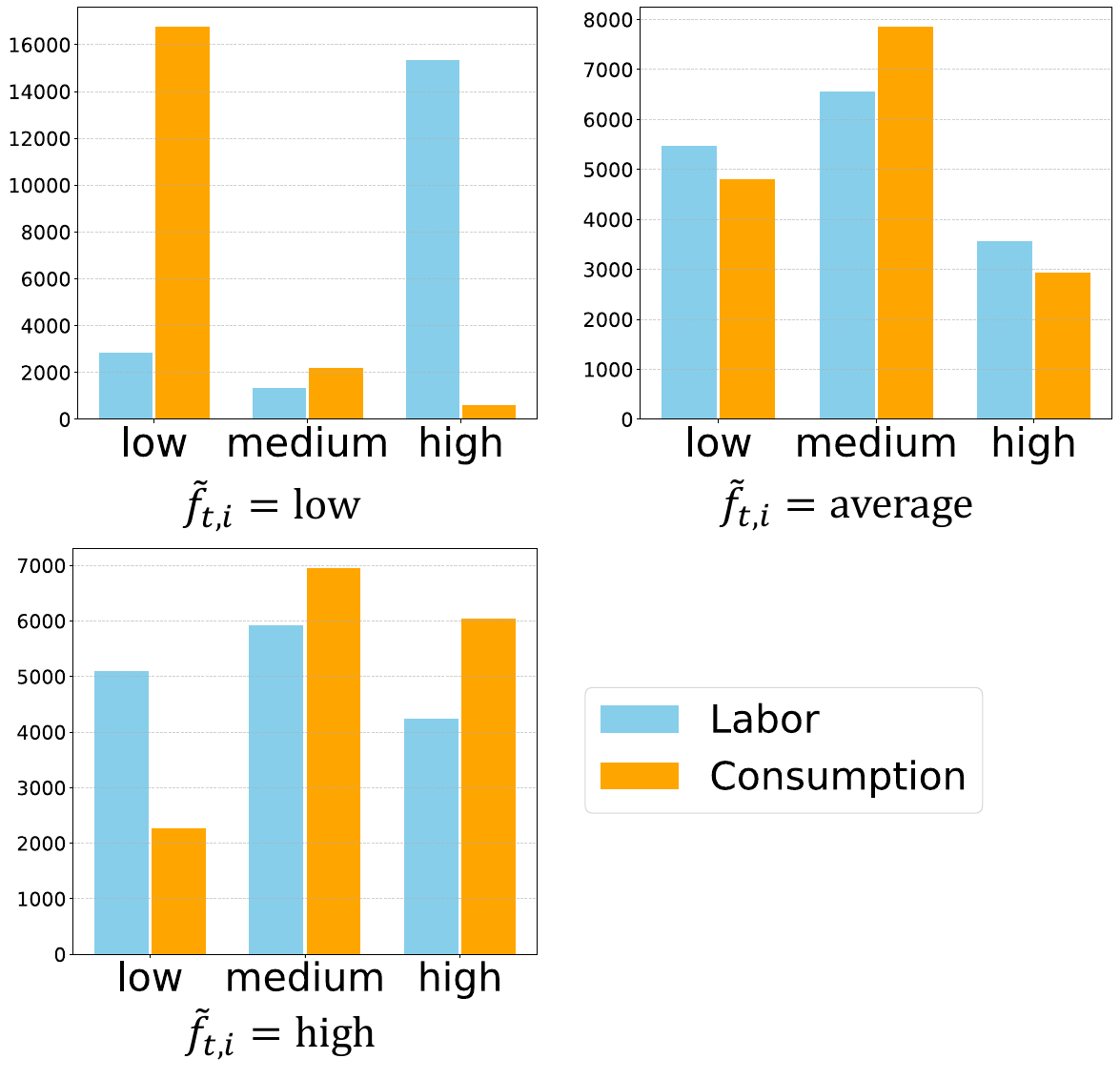}
  \caption{Household labor supply and consumption choices across economic conditions. Each panel reports the distribution of categorical actions ($a^L_{t,i}$ and $a^C_{t,i}$) for labor and consumption.}
  \label{Fig:action_bars}
\end{figure}

\subsection{RQ1: Behavioral Validation}
In RQ1, we conduct a sanity check to verify that LLM-based households are behaviorally grounded in the economic environment. This analysis examines whether agents’ economic actions and expressed opinions respond coherently and consistently to their experienced financial conditions, providing a necessary validation for subsequent trajectory- and population-level analyses.

Figure~\ref{Fig:action_bars} illustrates household labor supply and consumption choices, conditioned on relative income status. The distributions of categorical actions (low, medium, high) are aggregated across all simulation runs. As shown in Figure~\ref{Fig:action_bars}, low-income households predominantly select high labor supply and low consumption, whereas high-income households exhibit the opposite tendency, with lower labor supply and higher consumption. Middle-income households display intermediate behavior. These shifts across income groups are consistent with basic economic intuition under different budget constraints, indicating that agents adapt their economic behavior in response to their financial conditions. In Appendix~\ref{Sec:prompt_ablation}, we further assess the robustness of these observations by adjusting the prompt.

\begin{figure}[tb]
    \centering
    \begin{subfigure}[t]{0.48\linewidth}
        \centering
        \includegraphics[width=\linewidth]{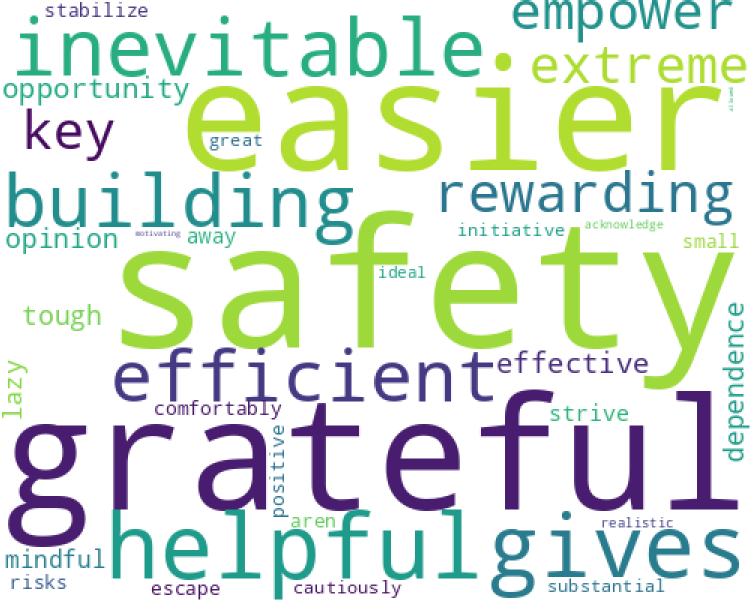}
        \caption{$h_{t,i}=(\text{POS},\text{low})$}
    \end{subfigure}
    \hfill
    \begin{subfigure}[t]{0.48\linewidth}
        \centering
        \includegraphics[width=\linewidth]{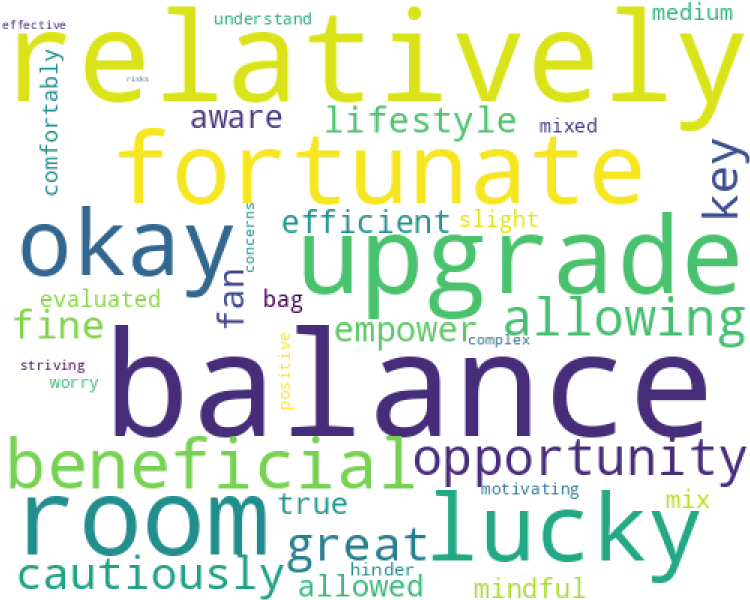}
        \caption{$h_{t,i}=(\text{POS},\text{high})$}
    \end{subfigure}
    \vspace{0.5em}
    \begin{subfigure}[t]{0.48\linewidth}
        \centering
        \includegraphics[width=\linewidth]{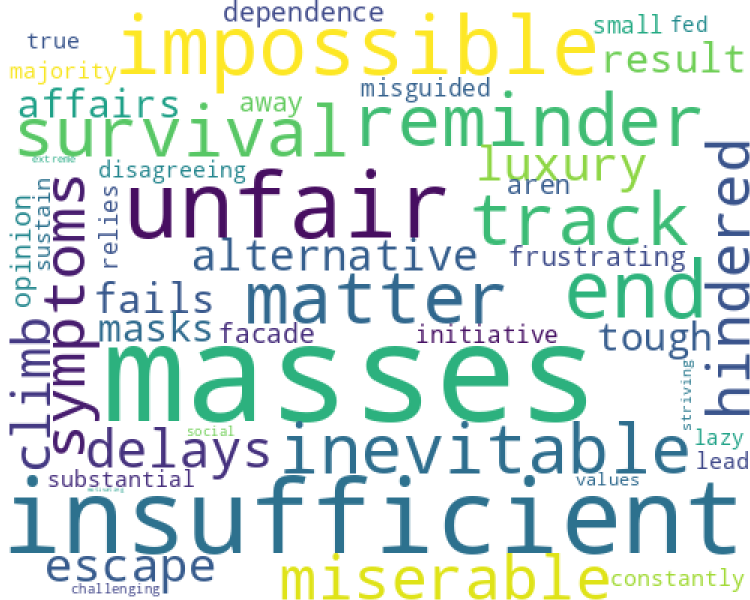}
        \caption{$h_{t,i}=(\text{NEG},\text{low})$}
    \end{subfigure}
    \begin{subfigure}[t]{0.48\linewidth}
        \centering
        \includegraphics[width=\linewidth]{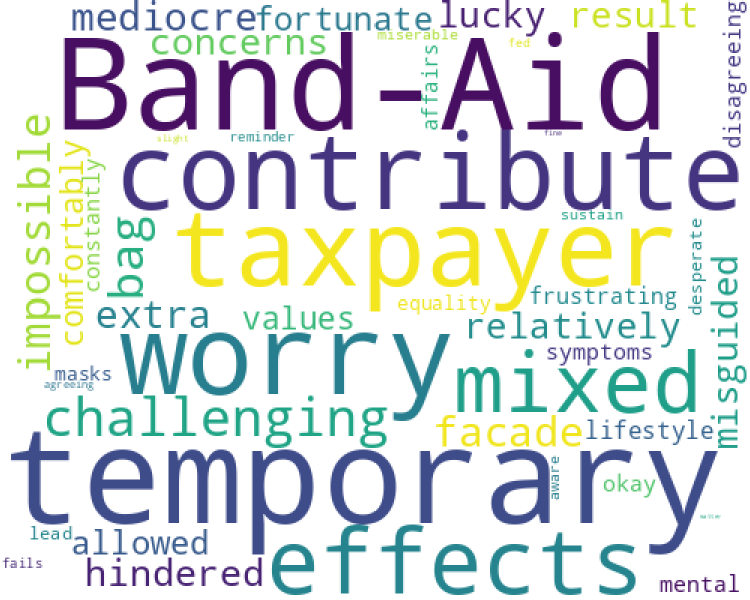}
        \caption{$h_{t,i}=(\text{NEG},\text{high})$}
    \end{subfigure}
    \caption{TF-IDF-based word clouds of household opinion texts across four internal states.}
    \label{Fig:wordclouds}
\end{figure}

\begin{figure*}[tbp]
    \centering
    \begin{subfigure}[t]{0.32\linewidth}
        \centering
        \includegraphics[width=\linewidth]{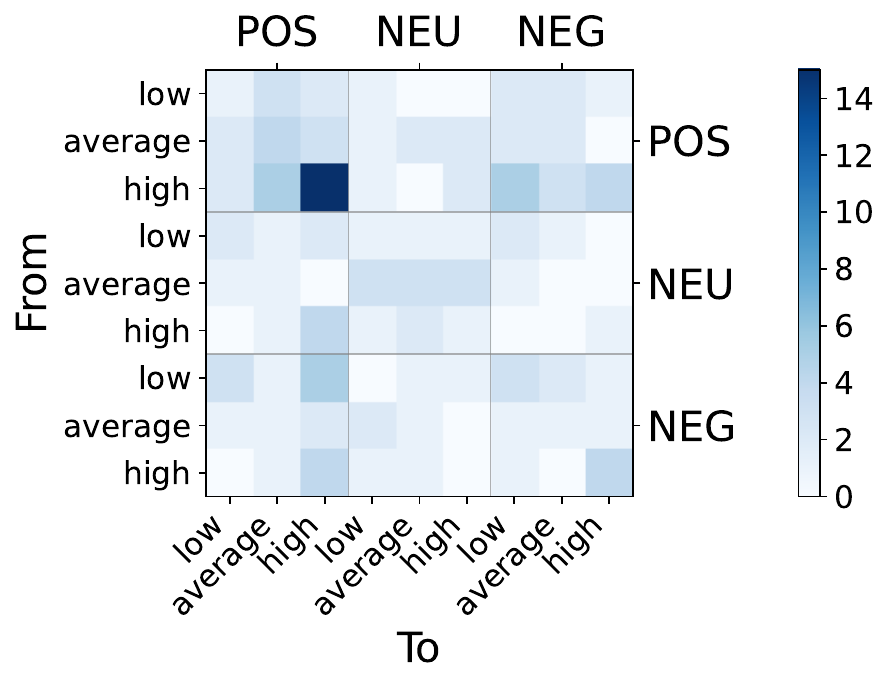}
        \caption{Cluster 1}
    \end{subfigure}
    \hfill
    \begin{subfigure}[t]{0.32\linewidth}
        \centering
        \includegraphics[width=\linewidth]{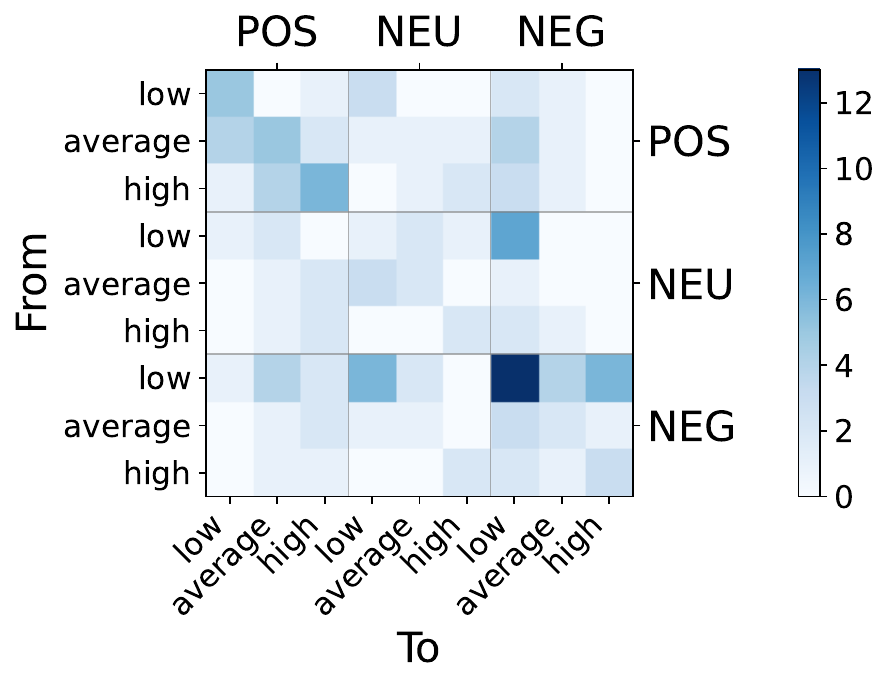}
        \caption{Cluster 2}
    \end{subfigure}
    \hfill
    \begin{subfigure}[t]{0.32\linewidth}
        \centering
        \includegraphics[width=\linewidth]{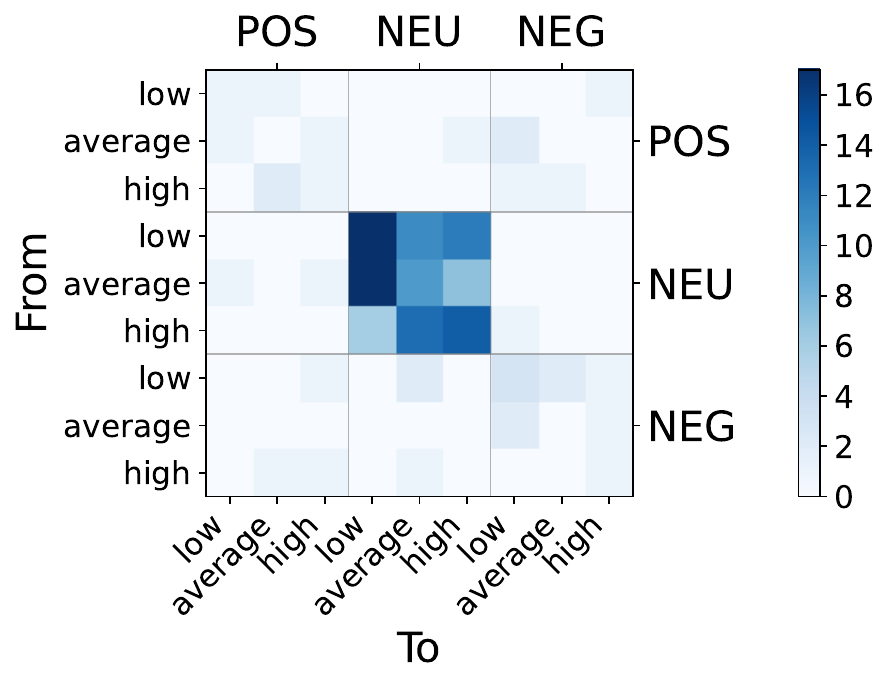}
        \caption{Cluster 3}
    \end{subfigure}
    \medskip
    \begin{subfigure}[t]{0.32\linewidth}
        \centering
        \includegraphics[width=\linewidth]{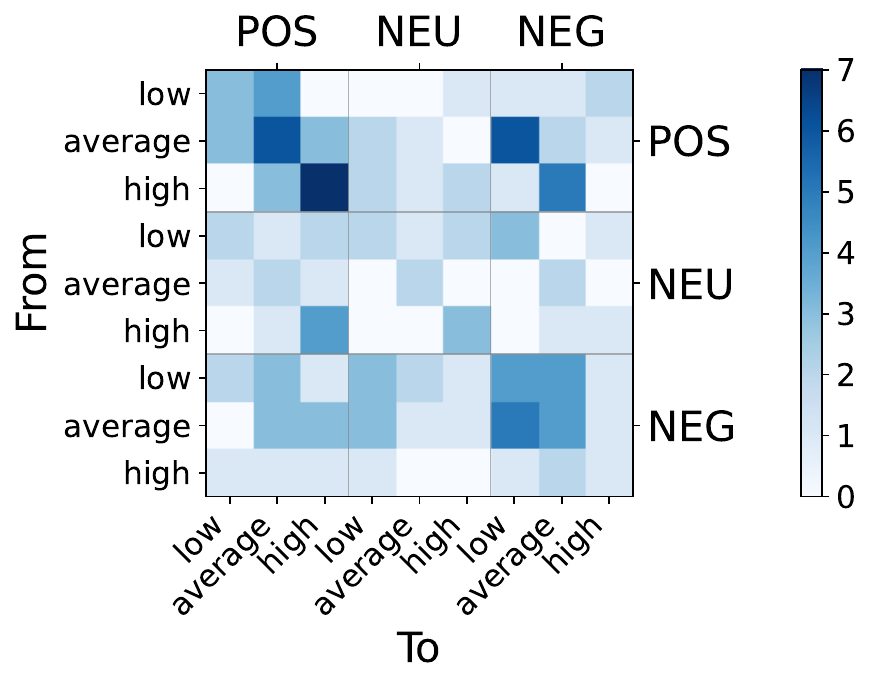}
        \caption{Cluster 4}
    \end{subfigure}
    \hfill
    \begin{subfigure}[t]{0.32\linewidth}
        \centering
        \includegraphics[width=\linewidth]{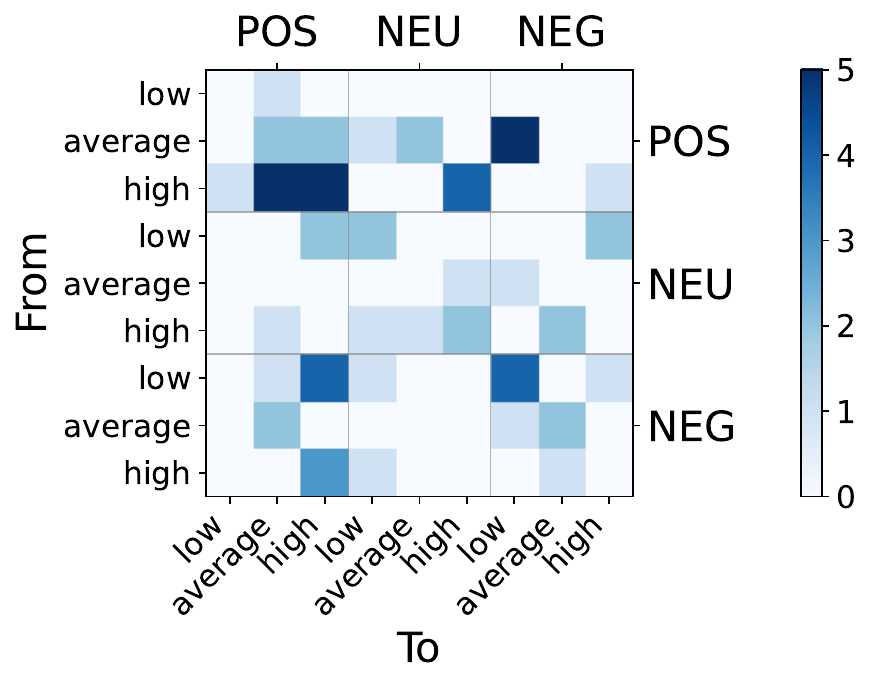}
        \caption{Cluster 5}
    \end{subfigure}
    \hfill
    \begin{subfigure}[t]{0.32\linewidth}
        \centering
        \includegraphics[width=\linewidth]{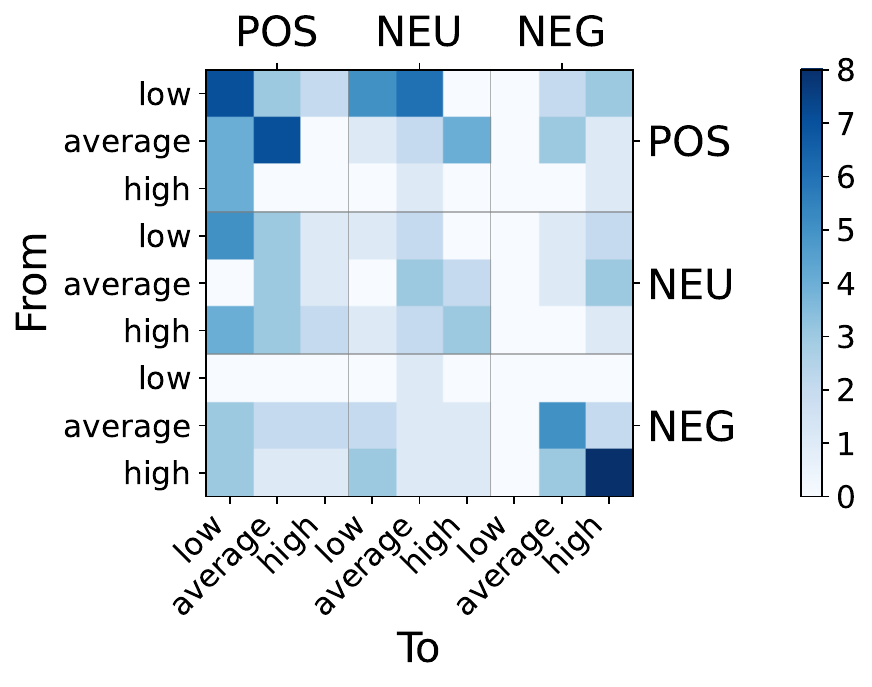}
        \caption{Cluster 6}
    \end{subfigure}
    \caption{Representative internal state transition patterns obtained by clustering transition count vector $\bm{n}_i$ across all simulations. Each panel visualizes the representative agent’s state transition matrix as a heatmap, where rows (From) indicate the internal state $h_{t-1,i}$, columns (To) indicate $h_{t,i}$, and color intensity denotes transition frequency.}
    \label{Fig:heatmap_transactions}
\end{figure*}

Figure~\ref{Fig:wordclouds} provides qualitative evidence on how households’ expressed opinions vary with their internal states. The TF-IDF~\citep{tf_idf}-based word clouds visualize salient terms extracted from household opinion texts under combinations of sentiment $\tilde{\psi}_{t,i}$ and income status $\tilde{f}_{t,i}$. For positive-sentiment households with low income, frequently appearing terms such as \textit{safety}, \textit{gratitude}, and \textit{helpful} suggest appreciation for the provided of a safety net. Positive-sentiment high-income households emphasize words including \textit{balance}, \textit{opportunity}, and \textit{upgrade}, indicating an understanding of the trade-off between institutional support and labor opportunities, alongside an opportunity-oriented perspective. In contrast, negative-sentiment low-income households focus on terms such as \textit{unfair}, \textit{insufficient}, and \textit{survival}, reflecting experienced hardship and heightened concerns about future insecurity. Negative-sentiment high-income households, characterized by words such as \textit{Band-Aid}, \textit{temporary}, and \textit{taxpayer}, express critical views toward existing institutions and concerns regarding redistribution and perceived burdens. The observed qualitative differentiation in salient terms across internal states indicates that opinion generation is coherently grounded in agents’ experienced economic situations, verifying cross-modal consistency between economic conditions, actions, and opinion content.

\subsection{RQ2: Individual OD}

In RQ2, we examine how individual opinions evolve in response to environmental experiences. 

\begin{figure}[tb]
  \centering
  \includegraphics[width=\linewidth]{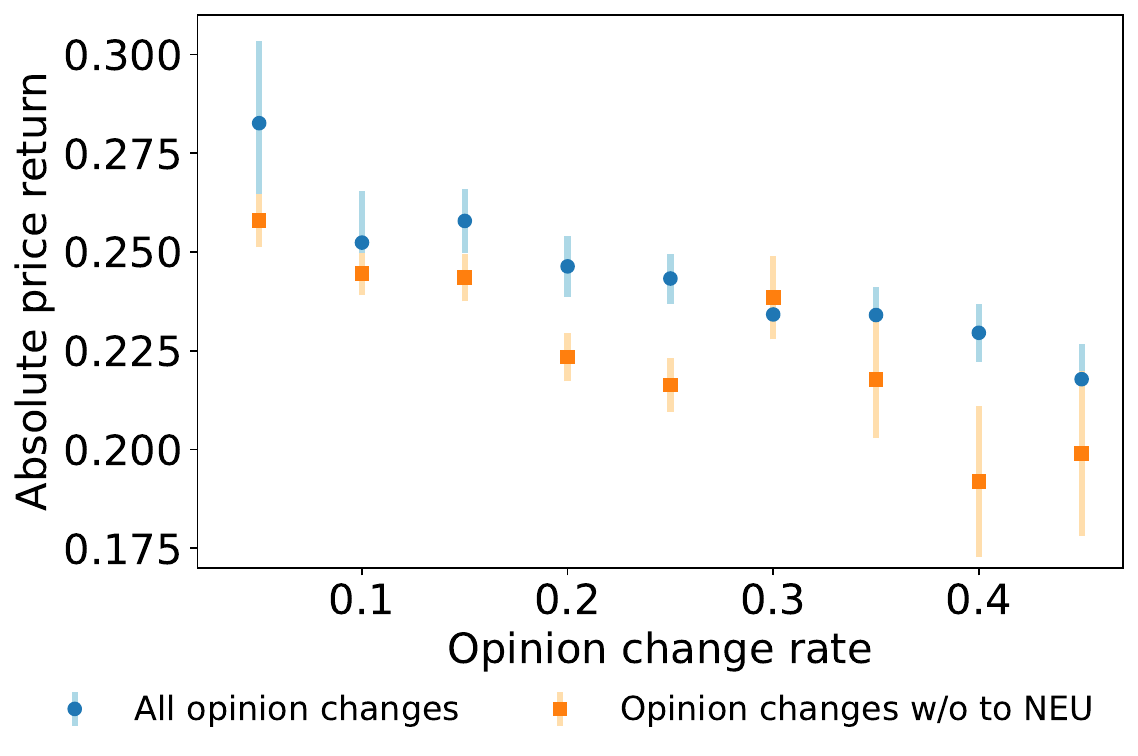}
  \caption{Relationship between different types of opinion change rates $\delta_t,\delta_t^{'}$ and price fluctuations. Absolute price return means the corresponding average absolute log price change $|\log p_{t,i}/p_{t-1,i}|$. Blue circles count all sentiment changes $\delta_t$, whereas orange squares exclude transitions to the neutral (NEU) class $\delta_t^{'}$. Only bins with more than 100 observations are shown. Error bars indicate the standard error of the mean (SEM).}
  \label{Fig:stance_change_vs_price_change}
\end{figure}

\begin{table*}[tbp]
    \centering
    \renewcommand{\arraystretch}{1.2}
    \setlength{\tabcolsep}{5pt}
    \begin{tabular}{rcccccc}
        \toprule
        & \multicolumn{4}{c}{\textbf{$G_t$ v.s.}} &
        \multicolumn{2}{c}{\textbf{$|\log (\bar{p}_t/\bar{p}_{t-1})|$ v.s.}} \\
        \cmidrule(lr){2-5} \cmidrule(lr){6-7}
        &$P_t^\Delta(0.2)$ &$P_t^\Delta(0.4)$ & $P_t^\Delta(0.6)$ & Std. of $\Psi_t$ & $W_1(\Psi_{t-1},\Psi_t)$ & $W_1(\mathcal{O}_{t-1},\mathcal{O}_t)$  \\
        \midrule

        Pearson 
        & \makecell{$0.149^{**}$ \\ {\footnotesize (0.016, 0.272)}} 
        & \makecell{$0.192^{***}$ \\ {\footnotesize (0.087, 0.280)}} 
        & \makecell{$0.222^{***}$ \\ {\footnotesize (0.120, 0.333)}} 
        & \makecell{$0.145^{***}$ \\ {\footnotesize (0.072, 0.234)}} 
        & \makecell{$0.145^{**}$ \\ {\footnotesize (0.009, 0.264)}} 
        & \makecell{$0.148^{**}$ \\ {\footnotesize (0.010, 0.267)}} \\

        Spearman
        & \makecell{$0.144^{**}$ \\ {\footnotesize (0.012, 0.271)}} 
        & \makecell{$0.181^{***}$ \\ {\footnotesize (0.085, 0.281)}} 
        & \makecell{$0.221^{***}$ \\ {\footnotesize (0.133, 0.358)}} 
        & \makecell{$0.155^{***}$ \\ {\footnotesize (0.073, 0.232)}} 
        & \makecell{$0.142^{**}$ \\ {\footnotesize (0.004, 0.202)}} 
        & \makecell{$0.118^{**}$ \\ {\footnotesize (0.002, 0.199)}} \\

        \bottomrule
    \end{tabular}
    \caption{Correlation between (i) economic inequality and collective opinion measures, and (ii) macroeconomic instability and opinion dynamics. 
    Each cell reports the correlation coefficient with the 95\% bootstrap confidence interval ($200$ resamples) shown below in parentheses. 
    To ensure independence and avoid pseudo-replication, bootstrap correlations are computed at the run level: for each of the $30$ independent simulation runs, we sample one time snapshot and evaluate correlations across these $30$ samples.
    \textit{Note.} * $\text{p}<0.10$, ** $\text{p}<0.05$, *** $\text{p}<0.01$.}
    \label{Tab:corrs}
\end{table*}

Figure~\ref{Fig:heatmap_transactions} visualizes representative patterns of individual internal state transitions from the KMeans clustering of state-transition count vectors $\{\bm{n}_i\}_{i=1}^n$ from entire set of simulations. For each cluster, we identify a representative agent whose state-transition count vector is closest to the corresponding cluster centroid in Euclidean distance. To assess clustering robustness, we repeated KMeans with 50 random seeds and obtained a mean Adjusted Rand Index (ARI) of $0.716 (\pm 0.073)$ across runs, indicating a stable clustering structure beyond random agreement. Figure~\ref{Fig:heatmap_transactions} provides evidence that OD is driven by individual, history-dependent responses to experienced economic feedback. Clusters 1 and 2 are characterized by transition patterns concentrated in (POS, high) and (NEG, low), indicating that both economic status and expressed opinions remain largely stable over time. Cluster 3 represents agents whose opinions remain neutral regardless of changes in economic status. In contrast, Clusters 4 and 5 display transitions between (POS, high) and (NEG, low), indicating that economic feedback drives opinion change. Cluster 6 shows that opinion changes are likewise driven by macroeconomic feedback, but through a qualitatively different transition structure involving (POS, low) and (NEG, high). Together, these observations indicate that LLM-based agents exhibit heterogeneous response patterns to macroeconomic feedback, which shape their individual opinion trajectories over time.

Figure~\ref{Fig:stance_change_vs_price_change} illustrates the relationship between sentiment change rates $\delta_t,\delta_t^{'}$ and contemporaneous price fluctuations. Figure~\ref{Fig:stance_change_vs_price_change} reveals a negative association between sentiment change rates and price fluctuations. Periods characterized by larger price changes tend to exhibit lower fractions of agents revising their sentiment, whereas higher rates of sentiment change are observed when price fluctuations are relatively modest. This pattern is robust to excluding transitions to neutrality, indicating that the effect is not driven by neutralization under normal conditions. This pattern suggests that agents’ sentiment orientations become more stable or reinforced under macroeconomic instability, indicating a form of rigidity in opinion formation. Such behavior is consistent with the threat–rigidity hypothesis~\citep{threat_rigidity}, which posits that under perceived threats, cognitive and behavioral responses tend to narrow.

\subsection{RQ3: Collective OD}

In RQ3, we shift our focus from individual-level OD to their collective manifestation at the population level. Building on the findings of RQ2, where we observed that a subset of individuals actively update their opinions in response to their economic feedback and opinion flexibility varies with macroeconomic conditions, we now examine how such individually driven dynamics shape the collective opinion distribution.

Table~\ref{Tab:corrs} shows correlations relating the Gini coefficient $G_t$ to two measures of opinion dispersion: the standard deviation of sentiment score distribution $\Psi_t$ at time $t$ and the polarization index $P_t^{\Delta}$. Across simulation runs, both the sentiment standard deviation and polarization index exhibit positive correlations with the Gini coefficient, which are statistically significant at the $5\%$ level based on Pearson correlation tests. These results indicate that widening economic disparities amplify not only the overall spread of opinions but also the degree of bimodal separation within the population. Importantly, this relationship mirrors empirical observations reported in the U.S., where increasing income inequality has been linked to heightened polarization~\citep{us_gini_pol}.

Table~\ref{Tab:corrs} also reports correlations between macroeconomic instability, measured by the absolute log return of the average price $|\log(\bar{p}_t/\bar{p}_{t-1})|$, and the magnitude of collective opinion change. Specifically, both the Wasserstein distance between consecutive sentiment distributions $W_1(\Psi_{t-1}, \Psi_t)$ and that between opinion embedding distributions $W_1(\mathcal{O}_{t-1},\mathcal{O}_t)$ exhibit positive and statistically significant correlations with price fluctuations. These results suggest that periods of heightened economic volatility are associated not merely with more dispersed opinions at a given time, but with larger reconfigurations of collective opinions over time.

Overall, our results indicate that widening economic inequality is associated with structural polarization, whereas macroeconomic volatility is linked to temporal instability in opinion distributions. Such relationships between macroeconomic dynamics and OD become observable only when LLM-based agents are modeled as decision-makers situated within, and directly affected by, the economic environments they experience.

\section{Related Work}
\paragraph{\textbf{OD Modeling}}
In studies of OD, a wide range of social interactions have been abstracted in order to theoretically characterize collective phenomena such as consensus formation, polarization, and the spread of misinformation. In the most fundamental models, including the DeGroot model~\citep{degroot_model} and Friedkin–Johnsen model~\citep{friedkin_johnsen_model}, the process by which an individual updates their opinion as a weighted average of others’ opinions is formalized, enabling analysis of how social influence and trust structures shape collective opinion formation. 

Subsequently, models such as the bounded confidence models~\citep{deffuant_weisbuch_model,hegselmann_krause_model}, threshold models~\citep{granovetter_threshold_model,watts_network_threshold_model}, and Axelrod’s model~\citep{axelrod_dissemination_of_culture} introduce constraints on interactions based on social distance, providing explanations for phenomena such as opinion convergence within homogeneous groups, emergence of echo chambers, and self-organization of polarization. Furthermore, research grounded in social impact theory~\citep{social_impact_theory} focus on psychological biases, modeling the influence of non-rational factors such as majority effects and conformity on opinion formation~\citep{dynamic_social_impact1,dynamic_social_impact2}.

Recent studies have explored OD using LLM-based agents. \citet{llm_fallacy} highlights a distinctive advantage of LLMs by demonstrating that agents both generate and are influenced by argumentative fallacies during discussions, thereby internalizing rhetorical and reasoning-related phenomena that are abstracted away in classical OD models. Also, recent studies report that populations of LLM-based agents are strongly influenced by framing and confirmation bias~\citep{bias_llm_opinions1}, and tend to converge toward equity-oriented consensus~\citep{bias_llm_opinions2}.

While these studies demonstrate the potential of LLMs to enrich OD through linguistically grounded interaction and cognitive bias, they largely retain a view of OD as a process driven by communication alone. In contrast, we emphasize the role of non-linguistic factors such as economic conditions, lived experiences, and structural constraints in shaping opinion. Building on this perspective, we introduce the agent–environment coupling into OD modeling by grounding opinion formation in agents’ actions, economic feedback, and evolving constraints, enabling the bottom-up emergence of macro-level opinion patterns.

\paragraph{\textbf{LLM-based Social Simulations}} Recent studies demonstrate that LLMs provide a foundation for social simulation by enabling psychologically grounded behavior, persona-driven heterogeneity, and text-based interaction at scale. For example, \citet{generative_agents} propose \textit{Generative Agents} that exhibit human-like daily behaviors and social interactions, demonstrating how language-based cognition support emergent social phenomena. 

In the economic domain, both macroeconomic and financial market simulations suggest that LLM-based agents reproduce aggregate economic patterns. LLM-empowered agents are shown to generate macroeconomic regularities \citep{econagent}, while similar emergent patterns are observed in LLM-based financial market simulations \citep{hashimoto_prima,hirano_prima}. Extending this direction, recent large-scale simulations aim to model complex societies composed of many interacting LLM-based agents. For instance, \citet{agentsociety} present a large-scale agent society to study collective behaviors and social dynamics, while \citet{socioverse} introduce a world model for social simulation that integrates LLM-based agents with data from millions of real-world users.




This work deepens prior attempts to construct social simulacra by reexamining the role of the environment in LLM-based social simulations. Rather than serving as a descriptive context, the environment in our framework provides feedback as future action constraints and updates aggregated macro-level order, thereby endogenously generating collective and temporal heterogeneity. As a result, socioeconomic couplings such as inequality-driven polarization and the co-movement of economic instability and opinion distribution volatility emerge bottom-up from individual actions.

\section{Conclusion}
In this work, we investigated OD of LLM-based agents by connecting them to an environment that requires action and generates consequential feedback. By enacting LLMs as economic agents in a macroeconomic system, we showed that opinions are not formed solely through linguistic interaction, but are actively driven by agents’ own actions and the experiences that follow from them. Our results reveal a structured coupling between environmental dynamics and OD. Interactions with an action-demanding environment drive how individual opinions are formed through lived economic experiences. Furthermore, the resulting collective OD evolve in close alignment with the dynamics of the environment itself, giving rise to population-level patterns such as inequality-linked polarization that are inaccessible in language-only settings.

\section*{Limitations}

\paragraph{\textbf{Simplified linguistic interaction mechanisms}} To isolate the effects of action-mediated environmental feedback on opinion formation, interactions among agents are intentionally simplified in the present framework. Agents are exposed to limited textual opinions, and information provided by institutions such as firms and the government is symmetric across households. Consequently, richer social processes---such as deliberative discussion, persuasion, networked communication, or voting---are not modeled. These mechanisms may interact in complex ways with experience-driven opinion change. Extending the framework to incorporate more structured and heterogeneous interaction designs remains an important direction for future work.

\paragraph{\textbf{Fixed institutional rules and exogenous policy design}} The macroeconomic environment is governed by fixed, rule-based institutions, including taxation and redistribution schemes that do not adapt in response to agents’ opinions. While this design choice enables a clean analysis of how environmental dynamics drive opinion formation, it precludes the study of institutional change driven by collective beliefs or political pressure. Allowing institutional rules to change over time would enable the study of how OD interact with dynamic policy environments.

Moreover, policy interventions are not systematically compared across counterfactual regimes (e.g., with versus without intervention). Extending the framework to support controlled policy variations across otherwise identical environments would enable quasi-experimental analyses in the spirit of difference-in-differences (DiD).

\paragraph{\textbf{Limited bidirectionality between opinions and actions}}
Although the framework explicitly captures how action-conditioned experiences drive OD, it does not yet realize fully bidirectional dynamics between opinions and the environment. In particular, opinions in the current setting do not directly influence agents’ subsequent economic decisions. We argue that achieving such bidirectionality crucially depends on the design of linguistic interaction mechanisms: in real societies, decision-making is shaped not only by individual experiences but also by the perceived opinion climate formed through social communication. Designing interaction structures in which socially expressed opinions systematically affect agents’ choices may enable co-evolutionary dynamics between collective opinions and environmental trajectories. Investigating such socially mediated feedback loops remains an important challenge for future research.

\bibliography{custom}

\appendix

\begin{table*}[tbp]
\centering
\begin{tabular}{llll}
\toprule
                            & \textbf{Parameter}& \textbf{Notation} & \textbf{Value} \\
\midrule
\multirow{2}{*}{Government} & Base tax rate     & $\bar{\tau}$      & $0.010$\\
                            & Basic income      & $b$               & $5.000$\\
\midrule
\multirow{9}{*}{Firm}       & Capital share     & $\alpha$          & $0.100$\\
                            & Depreciation rate & $d$               & $0.005$\\
                            & Productivity      & $A$               & $0.950$\\
                            & Initial price     & $\bar{p}_{1,1}$   & $1.000$\\
                            & Initial wage      & $\bar{w}_{1,1}$   & $1.500$\\
                            & Initial inventory & $K_{1,1}$         & $20.000$\\
                            & Max capital for production & $K_{\text{max}}$ & $50.000$\\
                            & Price elasticity  & $\eta_p$          & $0.200$\\
                            & Wage elasticity   & $\eta_w$          & $0.030$\\
\midrule
\multirow{7}{*}{Household}  & Initial cash      & $\forall i~ M_{1,i}$ & $U(0,100)$\\
                            & Low labor range   & $\mathcal{I}^L_{\text{low}}$ & $[4,5)$\\
                            & Medium labor range& $\mathcal{I}^L_{\text{medium}}$ & $[5,6)$\\
                            & High labor range  & $\mathcal{I}^L_{\text{high}}$ & $[6,7)$\\
                            & Low consumption range & $\mathcal{I}^C_{\text{low}}$ & $[5,6)$\\
                            & Medium consumption range & $\mathcal{I}^L_{\text{medium}}$ & $[6,7)$\\
                            & High consumption range & $\mathcal{I}^L_{\text{high}}$ & $[7,8)$\\
\bottomrule
\end{tabular}
\caption{Full list of hyperparameters in our simulation of OD with LLM-based households.}\label{Tab:hyperparameter_details}
\end{table*}

\section{Detailed Simulation Configurations}\label{Sec:hyperparameter}

Table~\ref{Tab:hyperparameter_details} summarizes the experimental setup of our simulation of OD with LLM-based households. Each simulation was conducted over $T=150$ time steps with $n=20$ agents. The household economic status category $f_{t,i}$ is determined by thresholding the normalized income $z_{t,i}$. Thresholds $-1.282$, $-0.524$, $0.524$, and $1.282$ correspond to the boundaries between the five categories: very low, below-average, around-average, above-average, and very high. These thresholds correspond to the upper 10th and 30th percentiles of a standard normal distribution.

\begin{algorithm}[t]
\caption{LLM-based OD simulation with macroeconomic environment.}
\label{Alg:simulation_core}
\KwIn{
$n$ household personas
}
\For{$i = 1$ \KwTo $n$}{
  Initialize financial state $M_{t=0,i}$\;
  Initialize opinion trajectory $\langle O_i \rangle = \{\}$\;
}
\For{$t = 1$ \KwTo $T$}{
  \For{$i = 1$ \KwTo $n$}{
    Compute relative financial status $f_{t,i}$ from $\{M_{t,j}\}_{j=1}^n$\;
    Construct prompt from persona, history, environment state, and peer opinions\;
    Query LLM with the prompt to obtain $(L_{t,i}, C_{t,i}, O_{t,i})$\;
    Update household income Eq.(\ref{Eq:household_income})\;
    Update firm state (price $\bar{p}$ and wage $\bar{w}$ of a unit of good and labor, and inventory $K_{t,i}$) Eq.(\ref{Eq:update_inventory},\ref{Eq:update_price},\ref{Eq:update_wage})\;
    Append $O_{t,i}$ to $\langle O_i \rangle$\;
  }
}
\end{algorithm}

Algorithm~\ref{Alg:simulation_core} shows the detailed procedure of a simulation. Below, we describe the structure of the prompt used in our experiments, introducing its components step by step. We first specify the agent’s role and the basic economic setting in which it operates.

\begin{tcolorbox}[promptbox={Prompt Example 1: Role and economic setting}]
\ttfamily\small
You are a household in an economy. In this economy, there is only one good. Your money can be used only to buy the good or to pay tax. You will be happy if you consume the good. In order to buy a good, you have to earn money in some ways. One way to earn money is to work.
\end{tcolorbox}

Next, we condition each agent on a persona that specifies basic demographic attributes and personality traits. Specifically, the attributes sex, age, marital status, education level, occupation, and city are assigned via stratified sampling. In addition to demographic attributes, each agent is endowed with personality traits based on the Big Five personality model~\citep{big_five}. For each of the five dimensions---extroversion, agreeableness, conscientiousness, neuroticism, and openness to experience---one of two opposing trait descriptors (e.g., extroverted vs. introverted) is randomly sampled and assigned to the agent.

\begin{tcolorbox}[promptbox={Prompt Example 2: Persona conditioning}]
\ttfamily\small
Your persona to role-play is as follows:
sex: Female, age: 29, ..., trait1: introverted, trait2: agreeable, ...
\end{tcolorbox}

We next present the macroeconomic environment observable to agents. Prices, wages, and policy variables are provided exogenously as components of the environmental state.

\begin{tcolorbox}[promptbox={Prompt Example 3: Public information}]
\ttfamily\small
You have the following information:
Firm Announcement:
Current price per a unit of goods: 1.64, Current wage per a unit of labor: 1.46 The price has been strongly increasing in the long term. Recently, the price has been moderately increasing in the short term. The wage has been stable in the long term. Recently, the wage has been slightly increasing in the short term.

Government Announcement:
We keep the current basic income policy in order to ensure a minimum standard of living and support households currently facing financial difficulties. Each household receives 5 unit of cash as a basic income. The funding for basic income is covered by taxes paid by households with very high income. Tax rate is set to 20.0 percent for households with very high income, and 10.0 percent for other households, while households with very low income are exempt from taxation.
\end{tcolorbox}

In our experiments, opinion exchange among agents is simplified as follows. At each stage $(t,i)$, agents are provided with the opinions expressed by other agents up to the two most recent preceding stages as part of the prompt.

\begin{tcolorbox}[promptbox={Prompt Example 4: Others’ opinions}]
\ttfamily\small
Other households' opinions:
- The current system is unfair to those who ...
- ...
\end{tcolorbox}

Next, we provide agents with information about their own past economic experiences and state transitions. This component constitutes the core of our framework, as it explicitly conditions agents on their lived experiences. By incorporating this information, expressed opinions are not merely reflections of the assigned persona, but are formed as outcomes of past actions and environmental feedback.

\begin{tcolorbox}[promptbox={Prompt Example 5: Individual economic history}]
\ttfamily\small
Your individual information:
Holding cash: 33.69 Latest paid tax: 3.19 Latest received subsidy: 5.00
You received subsidy larger than the tax you paid. You were a net recipient of subsidies last time.
Your financial status was initially below-average income, last time below-average income and is currently below-average income
Your cash holding is currently slightly increasing compared to the last time.
\end{tcolorbox}

In addition, agents are provided with their own previously expressed opinion from an earlier time step. This information serves as a short-term memory that introduces temporal continuity in opinion expression, allowing opinions to evolve in relation to past beliefs while remaining responsive to newly experienced economic conditions.

\begin{tcolorbox}[promptbox={Prompt Example 6: Own last opinion}]
\ttfamily\small
Your last opinion: This society seems to ...
\end{tcolorbox}

Finally, agents are asked to simultaneously choose their actions and express an opinion. Responses are strictly constrained to a predefined JSON format. This design ensures that non-linguistic actions (labor and consumption) and linguistic outputs (opinions) are generated jointly on the basis of the same underlying economic experiences.

\begin{tcolorbox}[promptbox={Prompt Example 7: Decision and opinion task}]
\ttfamily\small
Based on this information, please provide the following decisions. Take all information into account.
1. How much do you work? The more you work, the higher you earn money to buy goods. You only spend your time to work, no additional resource is required. In general, those who are financially struggling should work hard. Answer must be either {"low", "medium", "high"}.
2. How much do you consume goods? You will be happy if you consume more. Answer must be either {"low", "medium", "high"}.
3. Clearly state your opinion in your words about this society. Take your financial situation and your work-consumption decision made above into account. But you must not be ambiguous even if you are not sure. Try to offer your own distinct perspective.
Decide your responses in the following JSON format. Do not deviate from the format, and do not add any additional words to your response outside of the format. Here are the answer format.

{"work": "low"/"medium"/"high", "consume": "low"/"medium"/"high", "opinion": <str>, "stance": <int from -10 to 10>}
\end{tcolorbox}

\section{Computing Infrastructure}\label{Sec:computing_infrastructure}

The experiment was conducted on a workstation equipped with the following hardware and software configurations:

\begin{itemize}
\item \textbf{CPU}: AMD Ryzen 9 3950X 16-Core Processor (32 threads, base clock 3.5 GHz, max boost 4.76 GHz)
\item \textbf{GPU}: 2 × NVIDIA RTX 6000 Ada Generation (48 GB VRAM each)
\item \textbf{Memory}: 64 GB DDR4 RAM
\item \textbf{Operating System}: Ubuntu 22.04.5 LTS (Jammy Jellyfish)
\item \textbf{CUDA Toolkit}: 11.5
\item \textbf{NVIDIA Driver Version}: 560.35.03
\item \textbf{CUDA Runtime Version}: 12.6
\end{itemize}

The complete source code used for the experiment, along with a full list of Python library versions and environment specifications, is provided in the supplementary material.

\section{Prompt Ablation}\label{Sec:prompt_ablation}
We examine whether the behavioral differentiation observed in RQ1 is driven by explicit prompt instruction or by the economic feedback mechanism itself. To isolate the effect of this instruction, we construct an ablated prompt by removing the sentence: \textit{In general, those who are financially struggling should work hard.} The full prompt specification is provided in Appendix~\ref{Sec:hyperparameter}.

We rerun the simulation under identical conditions using the modified prompt, aggregating results over 30 independent runs. To quantify behavioral differentiation, we define the following normalized difference measure for labor supply:
\begin{align}
\Delta^L(\tilde{f})=\frac{N_{\text{high}}^L(\tilde{f})-N_{\text{low}}^L(\tilde{f})}{N_{\text{high}}^L(\tilde{f})+N_{\text{low}}^L(\tilde{f})}
\end{align}
where $\tilde{f}$ denotes a financial status defined in Eq. (\ref{Eq:tilde_f}). Here, $N_{\text{high}}^L(\tilde{f})$ is the total number of time steps across all agents and simulation runs in which agents in state $\tilde{f}$ choose the \textit{high labor} action, and $N_{\text{low}}^L(\tilde{f})$ is defined analogously for the low labor. Similarly, we define $\Delta^C(\tilde{f})$ for consumption decisions:
\begin{align}
\Delta^C(\tilde{f})=\frac{N_{\text{high}}^C(\tilde{f})-N_{\text{low}}^C(\tilde{f})}{N_{\text{high}}^C(\tilde{f})+N_{\text{low}}^C(\tilde{f})}
\end{align}
where $N_{\text{high}}^C(\tilde{f})$ and $N_{\text{low}}^C(\tilde{f})$ denote the counts of high (low) consumption actions, respectively.

Table~\ref{Tab:prompt_ablation} summarizes the results. The results show that the direction of behavioral differentiation remains unchanged across conditions. Low-income households continue to exhibit higher labor supply and lower consumption, while high-income households display the opposite pattern. Although the magnitude of differentiation decreases slightly (approximately 5–6\%), the qualitative structure is preserved. These findings indicate that the explicit instruction strengthens, but does not generate, the observed behavioral patterns. Instead, the differentiation primarily arises from the economic feedback mechanism embedded in the environment rather than from direct prompt compliance.

\begin{table}[tbp]
\centering
\small
\setlength{\tabcolsep}{4pt}
\begin{tabular}{lcccc}
\toprule
& $\Delta^L$(\text{low}) & $\Delta^C$(\text{low}) & $\Delta^L$(\text{high}) & $\Delta^C$(\text{high})\\
\midrule
Baseline & $0.682$ & $-0.931$ & $-0.086$ & $0.463$ \\
w/o inst. & $0.644$ & $-0.872$ & $-0.083$ & $0.437$ \\
\bottomrule
\end{tabular}
\caption{Behavioral differentiation under prompt ablation. Baseline refers to the main simulation setting described in the paper. The w/o inst. condition corresponds to a modified prompt in which the sentence \textit{In general, those who are financially struggling should work hard.} is removed.}
\label{Tab:prompt_ablation}
\end{table}

\end{document}